\documentclass[preliminary]{eptcs}
\providecommand{\event}{GCM 2021} %
\pdfoutput=1
\usepackage{underscore}           %

\usepackage{graphicx}
\usepackage{subfigure}
\graphicspath{./graphics/}
\usepackage[table, dvipsnames]{xcolor}
\usepackage{amsmath,amssymb}
\usepackage{mathtools}
\usepackage{wrapfig}
\usepackage{longtable}
\usepackage{booktabs}
\usepackage{pythonhighlight}
\usepackage{array}
\usepackage{pgffor}
\usepackage{fontawesome}
\usepackage{xcolor}

\usepackage{amsthm}
\usepackage{mathtools}

\theoremstyle{plain}

\newtheorem{theorem}{Theorem}

\theoremstyle{definition}
\newtheorem{definition}{Definition}
\newtheorem{example}{Example}

{\bfseries}{\itshape}

\makeatletter
\def\bstr{b}
\def\bfstr{bf}
\def\cstr{c}
\def\fstr{f}
\def\strLst{A,B,C,D,d,E,F,G,H,I,J,K,L,M,N,O,P,Q,R,S,T,U,V,W,X,Y,Z}

\newcommand{\MkB}[1]{\expandafter\def\csname\bstr#1\endcsname{\mathbb{#1}}}
  \@for\i:=\strLst\do{%
    \expandafter\MkB \i     }

\newcommand{\MkBF}[1]{\expandafter\def\csname\bfstr#1\endcsname{\mathbf{#1}}}
  \@for\i:=\strLst\do{%
    \expandafter\MkBF \i     }

\newcommand{\MkCal}[1]{\expandafter\def\csname\cstr#1\endcsname{\mathcal{#1}}}
  \@for\i:=\strLst\do{%
    \expandafter\MkCal \i     }

\newcommand{\MkFrak}[1]{\expandafter\def\csname\fstr#1\endcsname{\mathfrak{#1}}}
  \@for\i:=\strLst\do{%
    \expandafter\MkFrak \i     }
    
\makeatother

\newcommand{\LinAc}[1]{\overline{\mathsf{Lin}}(#1)}

\newcommand{\ac}[1]{\mathsf{#1}} %

\newcommand{\RMatchGT}[3]{\mathsf{M}^{{\text{\tiny $#1$}}}_{#2}(#3)}

\newcommand{\obj}[1]{\mathsf{obj}(#1)}

\newcommand{\mIO}{\mathop{\varnothing}}

\newcommand{\bra}[1]{\left\langle #1\right\vert}
\newcommand{\ket}[1]{\left\vert #1\right\rangle}
\newcommand{\braket}[2]{\left\langle \left. #1 \right\vert #2\right\rangle}

\renewcommand{\vec}[1]{\underline{#1}}

\newcommand{\cond}[1]{\mathsf{cond}(#1)}

\newcommand{\jcOp}[1]{\hat{\bO}(#1)}

\newcommand{\ti}[1]{%
 \ensuremath{\vcenter{\hbox{\includegraphics{diagrams/#1.pdf}}}}%
}

\colorlet{h1color}{blue!70!black} %
\colorlet{h2color}{orange!90!black} %
\colorlet{h3color}{blue!40!white} %
\colorlet{h4color}{green!40!black} %

\colorlet{darkblue}{blue!40!black} %
\definecolor{qOrange}{RGB}{214,153,92}

\title{Stochastic Graph Transformation\\ For Social Network Modeling}
\author{Nicolas Behr
\institute{Univ. de Paris, CNRS, IRIF\\Paris, France}
\and
Bello Shehu Bello
\institute{Dept. of Computer Science\\ Bayero Univ. Kano, Nigeria}
\and
Sebastian Ehmes
\institute{Real-Time Systems Lab\\TU Darmstadt, Germany}
\and
Reiko Heckel\thanks{Corresponding author email address: \texttt{rh122@leicester.ac.uk}
}
\institute{School of Informatics\\ 
Univ. of Leicester, UK}
}
\def\titlerunning{Stochastic Graph Transformation For Social Network Modeling}
\def\authorrunning{N. Behr, B.S. Bello, S. Ehmes, R. Heckel}
\begin{document}
\maketitle

\begin{abstract}
    Adaptive networks model social, physical, technical, or biological systems as attributed graphs evolving at the level of both their topology and data. They are naturally described by graph transformation, but the majority of authors take an approach inspired by the physical sciences, combining an informal description of the operations with programmed simulations, and systems of ODEs as the only abstract mathematical description. We show that we can capture a range of social network models, the so-called voter models, as stochastic attributed graph transformation systems, demonstrate the benefits of this representation and establish its relation to the non-standard probabilistic view adopted in the literature. We use the theory and tools of graph transformation to analyze and simulate the models and propose a new variant of a standard stochastic simulation algorithm to recreate the results observed.
\end{abstract}

\section{Introduction}

Modeling and analyzing the dynamics of social networks allows scientists to understand the impact of social interactions on areas as diverse as politics (opinion formation, spread of (dis)information), economic development, and health (spread of diseases, update of vaccines)~\cite{read2008dynamic,baumann2020modeling,DBLP:journals/tcss/BolzernCN20}. 
Much of the more foundational literature approach social network analysis from a perspective informed by statistical physics~\cite{newman2008physics}
using a combination of mathematical models (differential equations) and programmed simulation, both derived from an intuitive understanding of the operation of the network. This works well for static networks, where the structure is fixed and changes are to node or link attributes only, but in complex adaptive networks~\cite{Gross2008} the interconnectedness of structure and data evolution poses additional challenges.

Stochastic typed attributed graph transformation~\cite{HLM06FI} is an obvious choice to formalize and analyze complex adaptive networks. The formalism provides both tool support for simulation and analysis and an established theory to derive mathematical models from the same rule-based descriptions, thus replacing the informal, and sometimes vague, descriptions in natural language. A case in point are the various voter models~\cite{Durrett2012,zschaler2012early,klamser2017zealotry}
which describe opinion formation in a network of agents. The operations, as described in~\cite{Durrett2012} seem clear enough. 

\begin{quotation}
\dots we consider two opinions (called 0 and 1) \dots; and on each step, we pick a discordant edge $(x,y)$ at random \dots. With probability $1 - \alpha$, the voter at $x$ adopts the opinion of the voter at $y$. Otherwise (i.e., with probability $\alpha$), $x$ breaks its connection to $y$ and makes a new connection to a voter chosen at random from those that share its opinion. The process continues until there are no edges connecting voters that disagree.
\end{quotation}

Intuitively, this is a graph transformation system of four rules over undirected graphs whose nodes are attributed by 0 or 1, shown below using $\circ$ and $\bullet$, respectively. 
Referring to~\cite{bp2019-ext,nbSqPO2019,behrCRRC,BK2020} for further details, we utilise a ``right-to-left’’ convention for rewriting rules in order to accommodate a natural order of composition of the rule algebra representations.%

\[
\begin{matrix}\bullet \hphantom {-}\circ \\[-0.675em] \!\!\!\!\!\!\backslash \\[-0.675em]\bullet \end{matrix}\;\leftharpoonup \;\begin{matrix}\bullet \!\!-\!\!\circ \\[-0.25em]\bullet \end{matrix}
\qquad
\begin{matrix}\bullet \hphantom {-}\circ \\[-0.675em] \;\;\;/ \\[-0.675em]\circ \end{matrix}\;\leftharpoonup \;
\begin{matrix}\bullet \!\!-\!\!\circ \\[-0.25em]\circ \end{matrix}
\qquad
\bullet \!\!-\!\!\bullet \;\leftharpoonup \;\bullet \!\!-\!\!\circ
\qquad
\circ \!\!-\!\!\circ \;\leftharpoonup \;\bullet \!\!-\!\!\circ
\]

The first pair of rules models \emph{rewiring} where an agents disconnects from another one with a different opinion to form a connection with a third agent of the same opinion. In the second pair of rules, an agent connected to one with a different opinion \emph{adopts} the opinion of the other. 
We could now associate rates with these rules, resulting in a stochastic graph transformation system that allows simulation using the SimSG tool~\cite{DBLP:journals/jot/EhmesFS19} and the derivation of differential equations using the rule algebra approach~\cite{bp2019-ext,nbSqPO2019, behrCRRC,BK2020}.

However, on closer inspection we discover a number of discrepancies. First, the model in~\cite{Durrett2012} and related papers is probabilistic, but without time. This is an abstraction of real-world behavior in social networks, where time is not discrete and actions not round-based. Arguably, a continuous-time model is a better representation of this behavior. Semantically, a stochastic graph transformation system induces a continuous-time Markov Chain (CTMC) --- an established model with clear links to logics, model checking and simulation techniques, while the operational model behind \cite{Durrett2012} and others is left informal, but could be formalized as a discrete-time Markov chain (DTMC) or decision process (MDP)~\cite{bdg2016, bdg2019,bp2019-ext,BK2020}. %

More specifically, looking at the description of the operation, this is a two- or three-step process, where first a conflict edge is selected at random, then a decision made based on the fixed probability $\alpha$ between adopting another opinion, and rewiring which requires another random selection of a node with the same opinion. That means, in the rewiring case, the combined behavior of the operation is not reflected directly by the rewiring rules above, which choose all three nodes first. Also, the race between the rewiring and adopt rules in a stochastic graph transformation system depends on the number of available matches (the candidates for the 3rd agent to link) while the probability is fixed in the original formulation. Stochastic graph transformation realizes a mass action semantics which reflects the behavior of physical, chemical and  biological processes where the frequency of actions (formally the jump rate of the CTMC) depends on both a rate constant and the concentration or amount available of the input materials required for the action. 

In our view, the informal description of operations, lack of continuous time, and non-standard selection and execution procedure all represent weaknesses in the formulation which inhibit a natural mathematical interpretation of the operational behavior of the model and consequently a formal and systematic connection of this operational description with the simulation and equational model. 

Other authors have studied algorithmic problems in social networks, such as what subset of actors to target with interventions (e.g., in marketing or public health campaigns) so as to maximise the impact of the campaign on the network as a whole~\cite{DBLP:journals/toc/KempeKT15,DBLP:journals/kbs/LiuYWLLT16}. Such research is based on models of opinion formation and the spread of information in networks, typically divided into \emph{cascading} and \emph{threshold models}. In a cascading model, an actor is activated by a single message or activation from a peer. The behavior outlined above, where a discordant edge represents an activation, falls into this category. Instead, in a threshold model, an actor becomes active once a certain number of its peers are activated. 

Different models give rise to different centrality measure for actors~\cite{DBLP:journals/kbs/RiquelmeCMS18} and hence to different algorithmic solutions for discovering the most influential set of actors to target. These models assume static networks, but one could ask the question of centrality of actors for a dynamic network model given by a stochastic graph transformation system incorporating both the spread of information and the change of topology. 

In this paper, we present an approach addressing weaknesses in dynamic social network modelling by starting from a formal graph transformation model as operational description. We investigate the links to the existing voter models both analytically, using the rule algebra framework to establish how to translate operations and parameters to stochastic graph transformation, and experimentally trying to recreate the observed emergent behavior in the literature. Apart from making a convincing claim for the superiority of our methodological approach, this will allow us in the future to both compare new analysis results to existing ones in the literature.

We start by introducing the model as a stochastic graph transformation system in the format accepted by our simulation tool, then study the theory of relating the different CTMC and DTMC semantics to support different possible conversions between models before reporting on the experimental validation of the resulting models through simulation.

\section{Voter Models as Stochastic Typed Attributed Graph Rewrite Systems}\label{sec:voter-models}
\begin{figure}[h]
	\centering
	\includegraphics[width=0.50\textwidth]{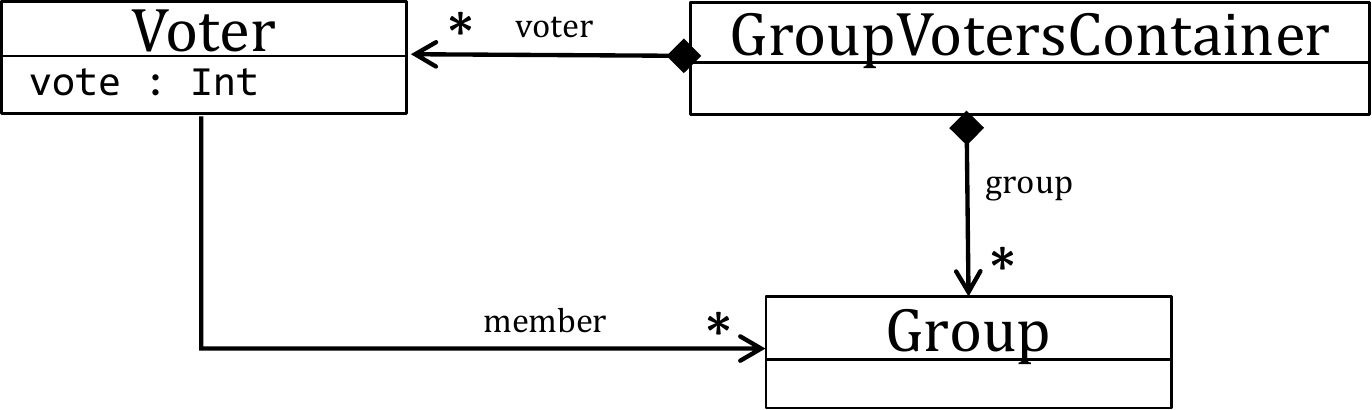}
	\caption{Type graph of the basic voter model}\label{fig:voter-model-tg}	
\end{figure}

We model voter networks as instances of the type graph below (Figure~\ref{fig:voter-model-tg}), where \emph{Group} nodes represent undirected links connecting \emph{Agent} nodes. This means that two agents are linked if they are  members of the same group. For now, the cardinality of each group is exactly 2, so these groups are indeed a model of undirected edges between agents (and  we are planning to generalize to groups with several members later).
We see the voter network as an undirected multi graph, i.e., parallel links between two voters $v1$ and $v2$ are permitted. 
That means, in our representation, $v1, v2$ can jointly make up one group $g1$ as well as another group $g2$. 
It is worth noting that the multi-graph interpretation is never explicitly stated in~\cite{Durrett2012} nor the literature building on it. 
However it is the only interpretation consistent with the usual assumptions made, in particular that rewiring is always possible after the selection of the 3rd node, which may or may not be connected to the first node $v1$ already, and that the total number of edges in the graph is constant, so we have to create a new edge on rewiring even if $v1$ and $v2$ are already connected.
\begin{figure}[h]
	\centering
	\includegraphics[width=0.80\textwidth]{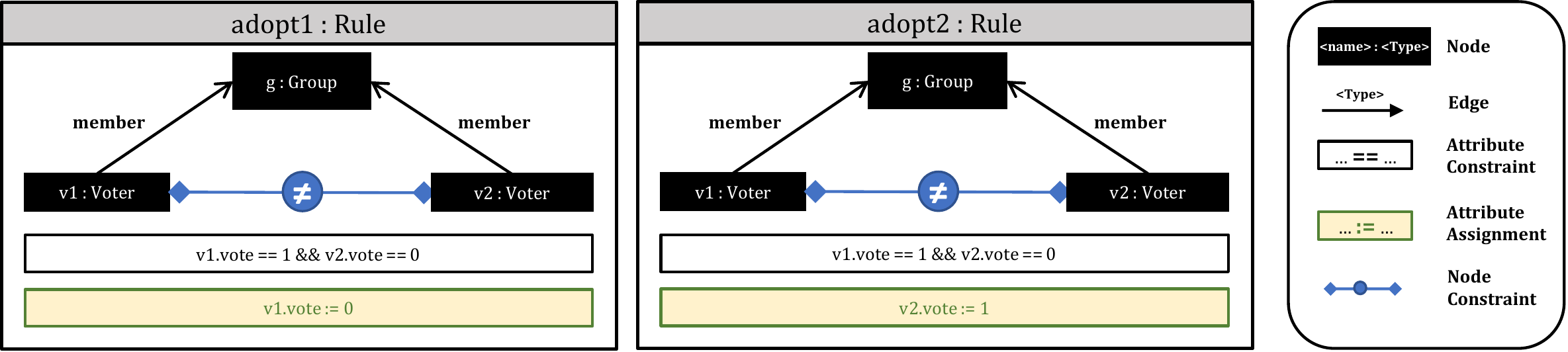}
	\includegraphics[width=0.80\textwidth]{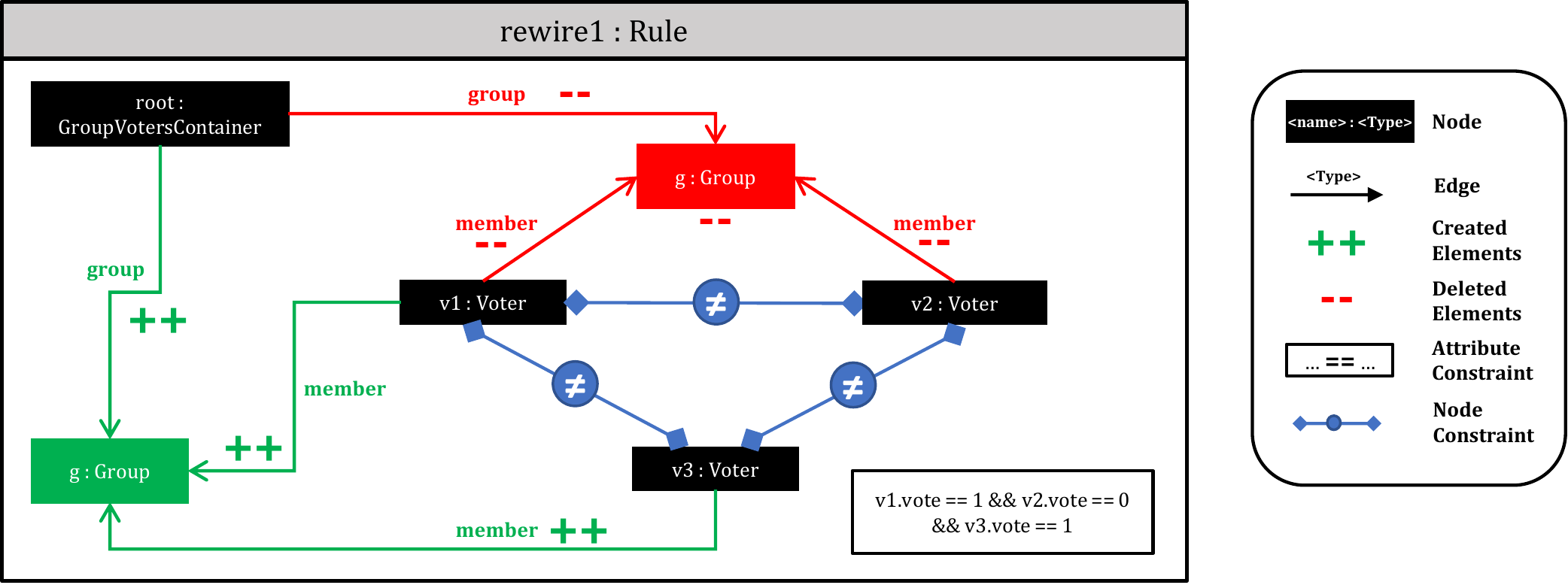}
	\includegraphics[width=0.80\textwidth]{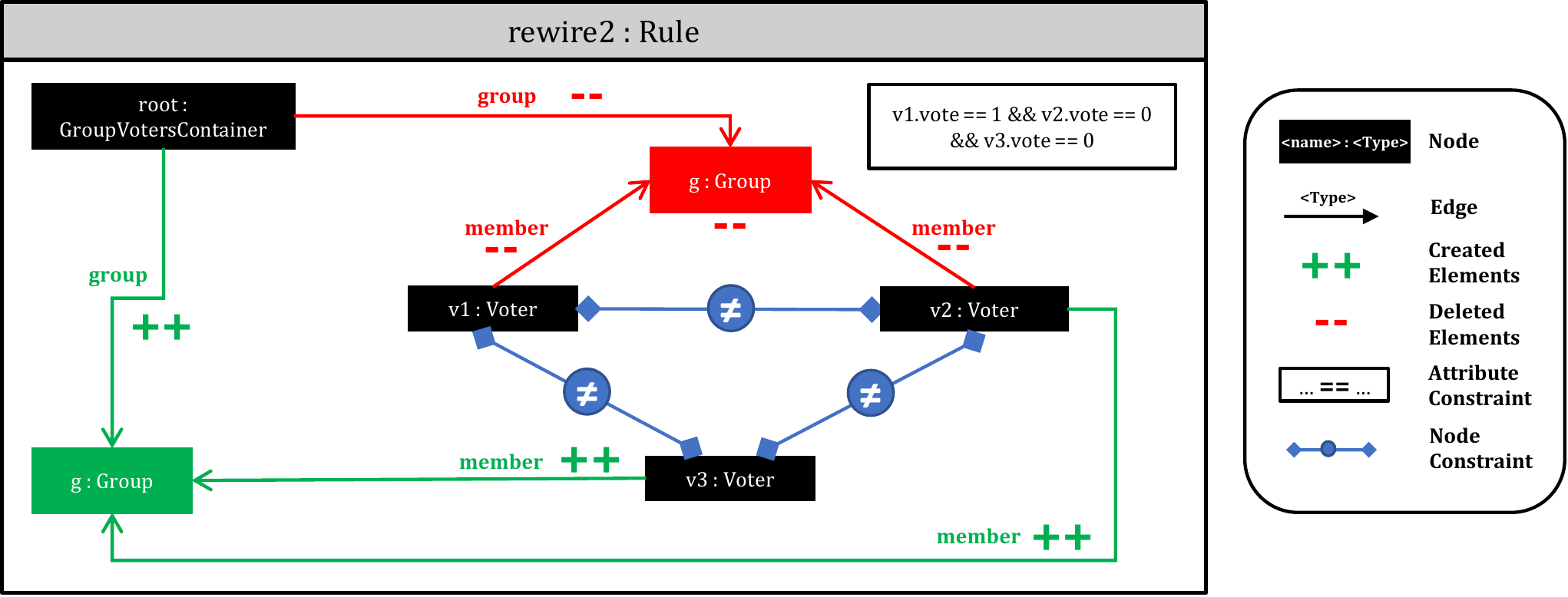}
	\caption{Rules of the basic voter model}\label{fig:voter-model-rules}
\end{figure}

In this paper we use SimSG \cite{DBLP:journals/jot/EhmesFS19} to simulate rule-based models, such as the basic voter model, for a given start graph and rates according to the semantics of stochastic graph transformations~\cite{HLM06FI}.
SimSG implements its own version of Gillespie's well-known algorithm~\cite{Gillespie1977} which describes the behavior of stochastic systems over time using a continuous-time Markov Chain with exponentially distributed transition delays.
SimSG is build upon the graph transformation rule interpreter eMoflon \cite{eMoflon}, which provides the means to define and execute rules. For example, there are four rules in the basic voter model, two dual variants each of the \emph{adopt} and \emph{rewire} operations. 
The graphs in Figure~\ref{fig:voter-model-rules} show a visual representation of these four rules as defined in the eMoflon::IBeX-GT\footnote{eMoflon::IBeX-GT: \url{https://emoflon.org/\#emoflonIbex}} syntax.
In these rules, black and red elements (\texttt{--}) specify context elements that need to be present in an instance graph, while green elements (\texttt{++}) will be created.
A pattern matcher will search for matches in an instance graph that fit these requirements.
As shown in Figure~\ref{fig:voter-model-rules}, in addition to structural constraints, eMoflon allows the specification of attribute conditions.
If a match is found, the rule can be applied by deleting all graph elements matching red elements in the rule, and creating new instances of all green elements.  
In addition to the structural constraints and attribute conditions shown in the example, eMoflon-GT also allows the definition of more complex conditions, such as negative application conditions that filter out matches connected to prohibited graph structures. 
Beyond the specification of rules, SimSG allows for the annotation of rules with rates, either through value literals or statically evaluated arithmetic expressions.
Most importantly, eMoflon provides SimSG with an interface to its underlying incremental graph pattern matching engines, such as Viatra\cite{VIATRA}, Democles\cite{DEMOCLES} or the recently developed HiPE\footnote{HiPE: \url{https://github.com/HiPE-DevOps/HiPE-Updatesite}}, all of which can be used to find matches for rules and track observable patterns during simulations.
During each simulation, SimSG tracks occurrence counts of these patterns and provide the user with the option to plot these counts over the simulation time or save them to a file.

\section{Theory: CTMCs and DTMCs via Rule-algebraic Methods}\label{sec:theory}

Approaches to social network modeling can be broadly classified by their underlying semantics. In this paper, we consider the two major classes consisting of \emph{discrete-time Markov chains (DTMCs)} and \emph{continuous-time Markov Chains (CTMCs)} (leaving for future work the class of Markov decision processes). We will utilize the \emph{rule algebra formalism}~\cite{bp2019-ext,nbSqPO2019, behrCRRC,BK2020} as our central technical tool, with CTMCs in so-called mass-action semantics implemented via the \emph{stochastic mechanics framework}~\cite{bdg2016, bdg2019,bp2019-ext, BK2020}. Here, the firing rate of a given rule for a system state at some time $t\geq 0$ is proportional to a \emph{base rate} (i.e., a positive real parameter) times the number of admissible matches of the rule into the system state.
Note however that even in the CTMC setting, this semantics is only one of several conceivable variants, and indeed we will study in this paper also a variation of rule-based CTMC semantics wherein rule activities in infinitesimal jumps are not weighted by their numbers of matches. In order to faithfully implement rule-based DTMCs, we will adopt the approach of~\cite{Behr2021}.

\subsection{Preliminaries: ``compositional'' rewriting theory}

We will focus throughout this paper on the case\footnote{Since the rules in our concrete models are \emph{linear} and will not involve vertex deletions nor creations, we could have equivalently opted for DPO- or SPO-semantics, which are well-known to coincide with SqPO-semantics in this special situation.} of Sesqui-Pushout (SqPO) semantics~\cite{Corradini2006}, which under certain conditions on the underlying base categories furnishes a ``compositional'' rewriting semantics~\cite{nbSqPO2019}, i.e., supports the requisite construction of rule algebras. Since the main theme of this paper is to advocate rewriting-based methods for implementing models, we discuss here the recent extension of the SqPO-formalism to include rules with conditions and under the influence of constraints~\cite{behrCRRC,BK2020}. We refer the interested readers to~\cite{BK2020} for the analogous formalism in the Double Pushout (DPO) setting.

Consider thus a \emph{base category} $\bfC$ that is \emph{$\cM$-adhesive} (for some class $\cM$ of monomorphisms), \emph{finitary}, possesses an $\cM$-initial object, $\cM$-effective unions and an epi-$\cM$-factorization, and such that all final pullback complements (FPCs) along composable pairs of $\cM$-morphisms exist (compare~\cite{BK2020}). Throughout this paper, we will consider the particular example of the category $\mathbf{uGraph}/T$ (for some $T\in \obj{\mathbf{uGraph}}$) of \emph{typed undirected multigraphs}, which according to~\cite{BK2020} satisfies all of the aforementioned requirements.

Let us briefly recall the salient points of SqPO-type ``compositional'' rewriting theory:

\begin{definition}[Rules]
Denote by $\LinAc{\bfC}$ the set of equivalence classes of linear rules with conditions,
\begin{equation}
\LinAc{\bfC}:= \{ R=(O\xleftarrow{o}K\xrightarrow{i}I,\ac{c}_I)\vert i,o\in \cM, \ac{c}_I\in \cond{\bfC}\}\diagup_{\sim}\,,
\end{equation}
where $R\sim R'$ if and only if $\ac{c}_I\dot{\equiv}\ac{c}_I'$ (i.e., if the conditions are equivalent for all matches of the rules) and the rules are isomorphic. The latter entails the existence of isomorphisms $\omega:O\xrightarrow{\cong} O'$, $\kappa:K\xrightarrow{\cong} K'$ and $\iota:I\xrightarrow{\cong} I'$ such that the obvious diagram commutes. We will adopt the convention to speak of ``a'' rule $R\in \LinAc{\bfC}$ to mean an equivalence class of rules.
\end{definition}

\begin{definition}
Given a rule $R\in \LinAc{\bfC}$ and an object $X\in \obj{\bfC}$, an \emph{SqPO-admissible match} $m\in \RMatchGT{}{R}{X}$ is an $\cM$-morphism $m:I\rightarrow X$ that satisfies the application condition $\ac{c}_I$. Then an \emph{SqPO-type direct derivation} of $X$ along $R$ with a match $m\in \RMatchGT{}{R}{X}$ is defined via a commutative diagram of the form
\begin{equation}
\ti{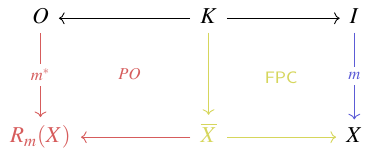}
\end{equation}
where the object $R_m(X)$ is defined up to the universal isomorphisms of FPCs and pushouts (POs).
\end{definition}
Note that in all of our applications, we will only consider objects up to isomorphisms, so we will in a slight abuse of notations speak of taking direct derivations of iso-classes $X\in \obj{\bfC}_{\cong}$ along (equivalence classes of) rules, understood as a mapping from iso-classes to iso-classes. With the focus of the present paper upon implementation strategies, we forgo here a complete review of the concepts of compositional rewriting theory (see e.g.\ \cite{behrCRRC,BK2020}), and limit ourselves to the following salient points. From hereon, $\bfC$ will always denote a category suitable for SqPO-type transformation.

\begin{definition}
Let $\cR_{\bfC}$ denote the \emph{$\bR$-vector space ``over rules''} spanned by basis vectors $\delta(R)$, where $\delta:\LinAc{\bfC}\xrightarrow{\cong} \mathsf{basis}(\cR_{\bfC})$ is an isomorphism. Let $\hat{\bfC}$ denote the \emph{$\bR$-vector space ``over states''}, which is defined via the isomorphism $\ket{.}:\obj{\bfC}_{\cong}\rightarrow \mathsf{basis}(\hat{\bfC})$ from isomorphism-classes of objects to basis vectors of $\hat{\bfC}$. We introduce the notation $End_{\bR}(\hat{\bfC})$ to denote the space of \emph{endomorphisms} on $\hat{\bfC}$ (i.e., of linear operators from $\hat{\bfC}$ to itself). Then we denote by $\rho_{\bfC}:\cR_{\bfC}\rightarrow End_{\bR}(\hat{\bfC})$ the morphism defined via
\begin{equation}\label{eq:defCanRep}
\forall R\in \LinAc{\bfC}, X\in \obj{\bfC}_{\cong}:\quad
\rho_{\bfC}(\delta(R))\ket{X} := \sum_{m\in \RMatchGT{}{R}{X}} \ket{R_m(X)}\,.
\end{equation}
The above definition extends to arbitrary elements of $\cR_{\bfC}$ and $\hat{\bfC}$ by \emph{linearity}, i.e., for $A=\sum_{j=1}^M \alpha_j\delta(R_j)$ and $\ket{\Psi}:=\sum_{X}\psi_X\ket{X}$, we define $\rho_{\bfC}(A)\ket{X}:= \sum_{j=1}^M\alpha_j\sum_{X} \psi_X \,\rho_{\bfC}(\delta(R_j))\ket{X}$.
\end{definition}

Intuitively, $\rho_{\bfC}$ takes a rule vector in $\cR_{\bfC}$ and delivers a transformation between state vectors, applying the rules to the input states, and ``weighing'' the resulting states according to the coefficients in the rule vector. This concept is useful in particular in order to formalize the jumps of a CTMC (see Section~\ref{sec:CTMC}).

\begin{theorem}[Cf.\ \cite{BK2020}]
\begin{itemize}
\item[(i)] The \emph{trivial rule} $R_{\mIO}:=[(\mIO\leftarrow\mIO\rightarrow\mIO,\ac{true})]_{\sim}\in \LinAc{\bfC}$ satisfies
\begin{equation}
\rho_{\bfC}(\delta(R_{\mIO}))=Id_{End_{\bR}(\hat{\bfC})}\,.
\end{equation}
\item[(ii)] Denote by $\bra{}:\hat{\bfC}\rightarrow \bR$ the so-called \emph{(dual) projection vector}, defined via $\braket{}{X}:=1_{\bR}$ for all $X\in \obj{\bfC}_{\cong}$. Extending this definition by linearity, $\bra{}$ thus implements the operation of \emph{summing coefficients}, i.e., for $\ket{\Psi}=\sum_{X}\psi_X\ket{X}$, we define $\braket{}{\Psi}:=\sum_X \psi_X\braket{}{X}=\sum_X\psi_X$. Then $\rho_{\bfC}$ satisfies the so-called \textbf{SqPO-type jump-closure property}:
\begin{equation}\label{eq:jc}
\forall R\in \LinAc{\bfC}:\quad \bra{}\rho_{\bfC}(\delta(R))
=\bra{}\jcOp{\delta(R)}\,,\quad  \jcOp{\delta(R)}:=\rho_{\bfC}(\delta([I\leftarrow I\rightarrow I,\ac{c}_I]_{\sim}))\,.
\end{equation}
\end{itemize}
\end{theorem}
The last point of the above theorem alludes to the important special case of rule-based linear operators that are \emph{diagonal} in the basis of $\hat{\bfC}$, since these implement operations of \emph{counting patterns}. In the particular case at hand, we may combine the definition of $\rho_{\bfC}$ in~\eqref{eq:defCanRep} with~\eqref{eq:jc} in order to obtain
\begin{equation}\label{ed:obs}
\bra{}\jcOp{\delta(R)}\ket{X}=\bra{}\rho_{\bfC}(\delta(R))\ket{X}=\sum_{m\in \RMatchGT{}{R}{X}}\braket{}{R_m(X)} =
\sum_{m\in \RMatchGT{}{R}{X}} 1_{\bR}= \vert \RMatchGT{}{R}{X}\vert\,.
\end{equation}
In this sense, $\jcOp{\delta(R)}$ as defined in~\eqref{eq:jc} ``counts'' the numbers of admissible matches of $R$ into objects.

\subsection{Rule-based CTMCs}\label{sec:CTMC}

Traditionally, CTMCs based upon transformation rules have been defined in so-called \emph{mass-action semantics}. In the rule algebra formalism, such types of CTMCs are compactly expressed as follows (with $\rho\equiv \rho_{\bfC}$):

\begin{theorem}[Mass-action semantics for CTMCs]\label{def:masCTMC}
Let $\cT:=\{(\kappa_j,R_j)\}_{j=1}^N$ denote a (finite) set of pairs of \emph{base rates} $\kappa_{j}\in \bR_{>0}$ and rules $R_j\in \LinAc{\bfC}$ (for $j=1,\dotsc,N$) over some category $\bfC$. Denote by $\mathsf{Prob}(\hat{\bfC})$ the space of (sub-)probability distributions over the state-space $\hat{\bfC}$, and choose an \emph{initial state} $\ket{\Psi_0}:=\sum_X \psi_X^{(0)}\ket{X}\in \mathsf{Prob}(\hat{\bfC})$. Then the data $(\cT,\ket{\Psi_0})$ defines a \textbf{SqPO-type rule-based CTMC} via the following \textbf{evolution equation} for the \textbf{system state} $\ket{\Psi(t)}:=\sum_X \psi_X(t)\ket{X}$ at time $t\geq0$:
\begin{equation}
\tfrac{d}{dt}\ket{\Psi(t)}=H\ket{\Psi(t)}\,,\quad \ket{\Psi(0)}=\ket{\Psi_0}\,,\quad H:=\rho(h)-\jcOp{h}\,,\quad h:=\sum_{j=1}^N \kappa_j\delta(R_j)
\end{equation}
Here, we used linearity to extend the definition of the jump-closure operator $\jcOp{.}$ of~\eqref{eq:jc} to arbitrary elements of $\cR_{\bfC}$, i.e., $\jcOp{h}:=\sum_{j=1}^N \kappa_j\jcOp{\delta(R_j)}$.
\end{theorem}

However, while mass-action semantics is of key importance in the modeling of chemical reaction systems, it is of course not the only conceivable semantics. In particular, as we shall demonstrate in the later part of this paper, for certain applications it will prove useful to utilize alternative adjustments of the ``firing rates'' of individual rules other than the one fixed in mass-action semantics.

\begin{definition}[Generalized rule-based CTMC semantics]\label{def:gsCTMC}
For a suitable base category $\bfC$, let $\cT:=\{(\gamma_j, R_j, W_j)\}_{j=1}^N$ be a set of triples of \emph{base rates} $\gamma_j\in \bR_{>0}$, \emph{rules} $R_j\in \LinAc{\bfC}$ and \emph{(inverse) weight functions} $W_j\in End(\hat{\bfC})_{diag}$ ($j=1,\dotsc,N$). Here, $End(\hat{\bfC})_{diag}$ denotes the space of \emph{diagonal operators} (with respect to the basis of $\hat{\bfC}$). Together with a choice of \emph{initial state} $\ket{\Psi_0}\in \mathsf{Prob}(\hat{\bfC})$, this data defines a \emph{rule-based SqPO-type CTMC} via the \emph{evolution equation}
\begin{equation}
\begin{aligned}
&\tfrac{d}{dt}\ket{\Psi(t)}=H_{\vec{W}}\ket{\Psi(t)}\,,\quad
\ket{\Psi(0)}=\ket{\Psi_0}\,,\quad
H_{\vec{W}}:= \sum_{j=1}^N \gamma_j\left(
\rho(\delta(R)_j)-\jcOp{\delta(R_j)}
\right)\frac{1}{W_j^{*}}\\
&\forall F\in End(\hat{\bfC})_{diag},\ket{X}\in \hat{\bfC}:\quad \frac{1}{F^{*}}\ket{X}:=\begin{cases}
\ket{X} \quad&\text{if }F\ket{X}=0_{\bR}\\
\frac{1}{\bra{}F\ket{X}}\ket{X} &\text{otherwise.}
\end{cases}
\end{aligned}
\end{equation}
\end{definition}

\begin{example}
Given a set of pairs of base rates and rules $\cT=\{(\kappa_j,R_j)\}_{j=1}^N$ for a  rule-based CTMC with mass-action semantics as in Definition~\ref{def:masCTMC}, one may define from this transition set in particular a \emph{uniformed} CTMC with infinitesimal generator $H_U$ defined as follows (with $\rho\equiv \rho_{\bfC}$):
\begin{equation}
H_U:=-Id_{End(\hat{\bfC})}+\sum_{j=1}^N p_j \cdot\rho(\delta(R_j))\frac{1}{\jcOp{\delta(R_j)}^{*}}\,,\quad p_j:=\frac{\kappa_j}{\sum_{j=1}^N \kappa_j}\,.
\end{equation}
Note in particular that since all base rates $\kappa_j$ are positive real numbers, the parameters $p_j$ are \emph{probabilities}. Moreover, applying $H_U$ to an arbitrary basis state $\ket{X}$, one finds that the overall jump rate (i.e., minus the coefficient of the diagonal part of $H_U$) is precisely equal to $1$, so that the $p_j$ in the non-diagonal part of $H_U$ encodes in fact the probability for the ``firing'' of rule $R_j$, regardless of the numbers of admissible matches of $R_j$ into the given state $\ket{X}$. Finally, under the assumption that both the mass-action and the uniformed CTMC admit a \emph{steady state}, we find by construction that
\begin{equation}
\lim\limits_{t\to\infty}\tfrac{d}{dt}\ket{\Psi(t)}=0\quad \Leftrightarrow \quad\lim\limits_{t\to\infty}H\ket{\Psi(t)}=0\quad 
\Leftrightarrow \quad \lim\limits_{t\to\infty}
H_U\left(\jcOp{h}\cdot \ket{\Psi(t)}\right)=0\,.
\end{equation}
Letting $\ket{\Psi_U(t)}$ denote the system state at time $t$ of the ``uniformed'' CTMC with generator $H_U$, and assuming $\ket{\Psi(0)}=\ket{\Psi_U(0)}$, the above entails in particular that $\ket{\Psi_U(t\to\infty)}= (\jcOp{h}^{*})^{-1}\cdot\ket{\Psi(t\to\infty)}$, i.e., the steady state of the CTMC generated by $H_U$ and the one of the CTMC generated by $H$ are related by an operator-valued rescaling (via the operator $(\jcOp{h}^{*})^{-1}$, and thus in general in non-constant fashion). A similar construction will play a key role when studying rule-based discrete-time Markov chains.
\end{example}

Another interesting class of examples is motivated via the ``Potsdam approach'' to probabilistic graph transformation systems as advocated in~\cite{giese2012}. 

\begin{example}[LCM-construction]\label{ex:LCM}
Given a generalized CTMC with infinitesimal generator $H_{\vec{W}}$ specified as in Definition~\ref{def:gsCTMC}, construct the least common multiple (LCM) $L$ of the inverse weight functions:
\begin{equation}
L:= LCM(W_1,\dotsc,W_N)\,.
\end{equation}
Then we define the \emph{LCM-variant} of the CTMC as
\begin{equation}
H_{LCM}:=H_{\vec{W}}\cdot L\,.
\end{equation}
By construction, $H_{LCM}$ does not contain any operator-valued inverse weights (in contrast to the original $H_{\vec{W}}$). Moreover, for certain choices of (inverse) weight functions $W_1, \dotsc, W_N$, the LCM-construction results in an infinitesimal generator $H_{LCM}$ in which all contributing rules have the same input motif $I$, thus making contact with the methodology of~\cite{giese2012}. 
\end{example}

\subsection{Rule-based DTMCs}\label{sec:dtmc}

Despite their numerous applications in many different research fields, to date discrete-time Markov chains (DTMCs) that are based upon notions of probabilistic transformation systems have not been considered in quite a comparable detail as the CTMC constructions. Following~\cite{Behr2021}, we present here a possible general construction of rule-based DTMC via a rule-algebraic approach:

\begin{definition}\label{def:DTMC}
For a suitable category $\bfC$, let $\cT=\{(\gamma_j,R_j,W_j)\}_{j=1}^N$ be a (finite) set of triples of positive coefficients $\gamma_j\in \bR_{>0}$, transformation rules $R_j\in \LinAc{\bfC}$ and (inverse) weight functions $W_j\in End(\hat{\bfC})_{diag}$ ($j=1,\dotsc,N$), with the additional constraint that
\begin{equation}
\sum_{j=1}^{N} \gamma_j\jcOp{\delta(R_j)}\cdot\frac{1}{W_j^{*}} = Id_{End(\hat{\bfC})}\,.
\end{equation}
Then together with an initial state $\ket{\Phi_0}\in \mathsf{Prob}(\hat{\bfC})$, this data defines a \emph{SqPO-type rule-based discrete-time Markov chain (DTMC)}, whose $n$-th state  (for non-negative integer values of $n$) is given by
\begin{equation}
\ket{\Phi_n}:=D^n \ket{\Phi_0}\,,\quad D:=\sum_{j=1}^N\gamma_j\rho(\delta(R_j))\cdot\frac{1}{W_j^{*}}\qquad (\rho\equiv\rho_{\bfC})\,.
\end{equation}
\end{definition}

\begin{figure}
\centering
\[
\ti{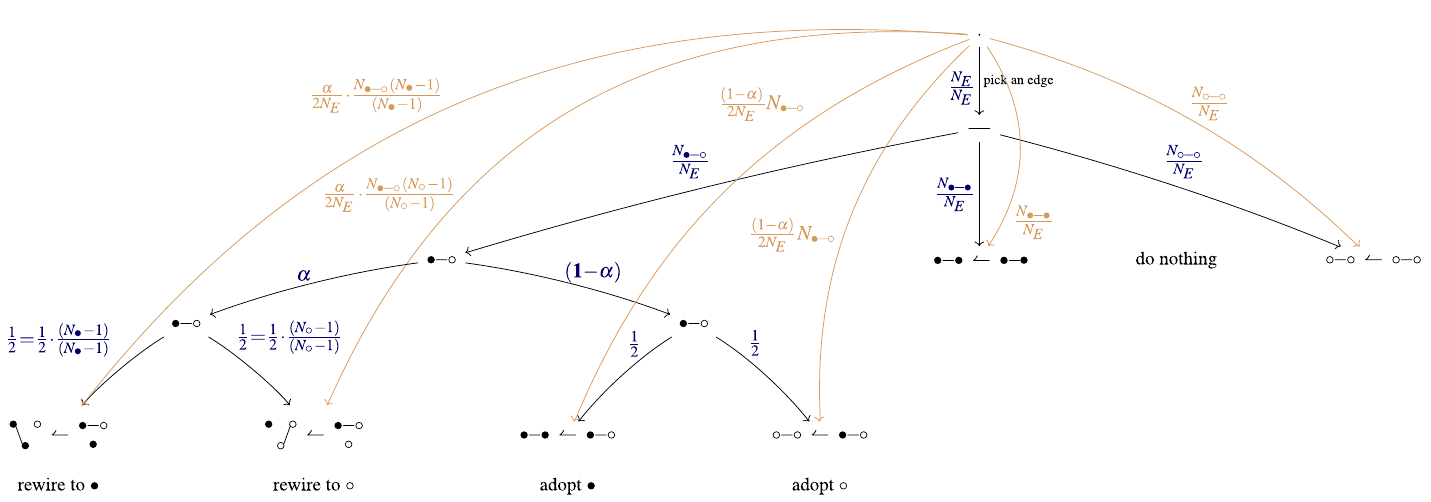}
\]
\caption{Specification of the adaptive voter model (AVM) according to~\cite{Durrett2012} via a probabilistic decision tree (black arrows), and as combined one-step probabilistic transitions ({\color{qOrange}orange arrows}).}\label{fig:AVM}
\end{figure}

\begin{example}[Adaptive Voter Model] As a typical example of a social network model, consider the specification of the \emph{adaptive voter model (AVM)} in the variant according to Durrett et al. [ref], which has as its \emph{input parameters} an \emph{initial graph state} $\ket{\Phi_0}=\ket{X_0}$ (with $N^{(0)}_{\bullet}$ and $N^{(0)}_{\circ}$ vertices of types $\bullet$ and $\circ$, respectively, and with $N_E$ undirected edges) and a \emph{probability} $0\leq \alpha\leq1$, and whose transitions as depicted in Figure~\ref{fig:AVM} are given via a form of a probabilistic decision procedure in several steps (black arrows): in each round of the AVM, an edge is chosen at random; if the edge is linking two nodes of different kind, with probability $p$, one of the rewiring rules is ``fired'', or with probability $(1-p)$ one of the adopt rules, respectively; otherwise, i.e., if the chosen edge is linking two vertices of the same kind, no action is performed. As annotated in {\color{darkblue}dark blue}, each phase of this probabilistic decision procedure is dressed with a probability (and so that all probabilities for a given phase sum to $1$). These probabilities in turn depend upon constant parameters (here, $p$ and $N_E$), as well as on \emph{pattern counts} $N_{\bullet}$, $N_{\circ}$, $N_{\bullet\!-\!\bullet}$, $N_{\bullet\!-\!\circ}$ and $N_{\circ\!-\!\circ}$ that dynamically depend on the current system state. Note that $N_V:=N_{\bullet}+N_{\circ}$ and $N_E=N_{\bullet\!-\!\bullet}+N_{\bullet\!-\!\circ}+N_{\circ\!-\!\circ}$ are \emph{conserved quantities} in this model, since none of the transitions create or delete vertices, but at most exchange vertex types\footnote{Strictly speaking, we implement the two vertex types $\bullet$ and $\circ$ as attributes in our \texttt{SimSG} implementation, which in the theoretical setting can be emulated via using self-loops of two different types; evidently, this amounts merely to a slight modification of the type graph plus the enforcement of some structural constraints (compare~\cite{BK2020}), thus for notational simplicity, we do not make this technical detail explicit in our diagrams.}, and since the transitions manifestly preserve the overall number of edges.

Upon closer inspection, the transition probabilities may be written in a form that permits to compare the ``firing'' semantics to the mass-action and to the generalized semantics used in rule-based CTMCs. According to~\eqref{ed:obs}, we may implement the operation of counting patterns via linear operators based upon ``identity rules'', since for arbitrary graph patterns $P$ and graph states $\ket{X}$,
\begin{equation}
N_P(X) = \bra{}\hat{O}_P\ket{X}\,,\quad \hat{O}_P:=\rho(\delta([P\leftarrow P\rightarrow P,\ac{true}]_{\sim}))\,.
\end{equation}
Since rule algebra theory is not the main topic of the present paper, we provide below some operator relations without derivation\footnote{In fact, for the simple case at hand, one may derive the relations heuristically, noting that applying $\hat{O}_{\bullet\!-\!\circ}\hat{O}_{x}$ to some graph state amounts to first counting all vertices of type $x$ (for $x\in \{\bullet,\circ\}$), followed by counting all $\bullet\!\!-\!\!\circ$ edges; performing this operation in one step, this amounts to either counting the $x$-type vertices separately (i.e., counting the pattern $\bullet\!\!-\!\!\circ\; x$), or counting the $x$-type vertices on the same location as counting the $\bullet\!\!-\!\!\circ$ patterns, thus explaining the two contributions in~\eqref{eq:auxObsAVM}.}, noting that they may be computed utilizing the fact that $\rho$ is a so-called representation of the SqPO-type rule algebra (cf.\ \cite[Thm.~3]{BK2020}):
\begin{equation}\label{eq:auxObsAVM}
\hat{O}_{\bullet\!-\!\circ}\hat{O}_{x}=\hat{O}_{\bullet\!-\!\circ\;x}+\hat{O}_{\bullet\!-\!\circ} \quad \Leftrightarrow\quad 
\hat{O}_{\bullet\!-\!\circ\;x}=\hat{O}_{\bullet\!-\!\circ}(\hat{O}_{x}-1)\qquad (x\in \{\bullet,\circ\})\,,
\end{equation}
which permit us to interpret the AVM model in the following way as a rule-based DTMC: starting by computing the overall one-step transitions and their respective probabilities ({\color{qOrange}orange arrows} in Figure~\ref{fig:AVM}), and using the above operator relations, we  find the rule-based DTMC generator
\begin{subequations}
\begin{align}
D_{AVM}&={\color{qOrange}\frac{\alpha}{2N_E}}
\rho\left(\delta\left(\begin{matrix}\bullet \hphantom {-}\circ \\[-0.675em] \!\!\!\!\!\!\backslash \\[-0.675em]\bullet \end{matrix}\;\leftharpoonup \;\begin{matrix}\bullet \!\!-\!\!\circ \\[-0.25em]\bullet \end{matrix}\right)\right){\color{qOrange}\frac{1}{(\hat{O}_{\bullet}-1)^{*}}}
+{\color{qOrange}\frac{\alpha}{2N_E}}\rho\left(\delta\left(\begin{matrix}\bullet \hphantom {-}\circ \\[-0.675em] \;\;\;/ \\[-0.675em]\circ \end{matrix}\;\leftharpoonup \;\begin{matrix}\bullet \!\!-\!\!\circ \\[-0.25em]\circ \end{matrix}\right)\right){\color{qOrange}\frac{1}{(\hat{O}_{\circ}-1)^{*}}}\label{eq:D-AVM-rewire}\\
&\quad+{\color{qOrange}\frac{(1-\alpha)}{2N_E}}\rho\left(\delta\left(
\bullet \!\!-\!\!\bullet \;\leftharpoonup \;\bullet \!\!-\!\!\circ
\right)\right)+{\color{qOrange}\frac{(1-\alpha)}{2N_E}}\rho\left(\delta\left(
\circ \!\!-\!\!\circ \;\leftharpoonup \;\bullet \!\!-\!\!\circ
\right)\right) \label{eq:D-AVM-adopt}\\
&\quad+{\color{qOrange}\frac{1}{N_E}}\rho\left(\delta\left(
\bullet \!\!-\!\!\bullet \;\leftharpoonup \;\bullet \!\!-\!\!\bullet
\right)\right)+{\color{qOrange}\frac{1}{N_E}}\rho\left(\delta\left(
\circ \!\!-\!\!\circ \;\leftharpoonup \;\circ \!\!-\!\!\circ
\right)\right) \label{eq:D-AVM-inactive}\,.
\end{align}
\end{subequations}
Interestingly, the particular semantics for the ``firing'' of transitions as encoding in this DTMC generator is neither a variant of mass-action semantics (wherein each rule would fire with a rate dependent on its number of matches, and so that the overall firing rate of all rules would be utilized to normalize all state-dependent firing rates to probabilities as described in Definition~\ref{def:DTMC}), nor do all transitions have state-independent, i.e., constant rates. Instead, the \emph{rewiring rules}~\eqref{eq:D-AVM-rewire} have rates proportional to $N_{\bullet\!-\!\circ}$; according to~\eqref{eq:auxObsAVM}, we have that $N_{\bullet\!-\!\circ}=N_{\bullet\!-\!\circ\;x}/(N_X-1)^{*}$ (for $x\in \{\bullet,\circ\}$), whence the firing rates of the rewiring transitions of the AVM model are found to be proportional to a state-dependent fraction of their mass-action rates! In contrast, the \emph{adoption} transitions~\eqref{eq:D-AVM-adopt} as well as the \emph{inactive} transitions~\eqref{eq:D-AVM-inactive} are found to follow standard mass-action semantics, which might raise the question of whether the overall ``mixed'' semantics chosen for the AVM model indeed reflect the intended intuitions and empirical findings.

Finally, in view of simulation experiments, we would like to advocate the use of dedicated and high-performance Gillespie-style stochastic graph transformation software such as in particular \texttt{SimSG}~\cite{DBLP:journals/jot/EhmesFS19} (cf.\ Section~\ref{sec:simulations}). More concretely, one may take advantage of the rule-algebraic formulation of DTMCs and CTMCs to identify for a given DTMC with generator $D$ a particular CTMC based upon the \emph{uniform} infinitesimal generator $H_U:=D-Id_{End(\hat{\bfC})}$, so that (if both limits exist) the two alternative probabilistic systems have the same limit distribution (i.e., $\ket{\Phi_{\infty}}=\lim\limits_{t\to\infty}\ket{\Psi(t)}$):
\begin{equation}
D\ket{\Phi_{\infty}}=\ket{\Phi_{\infty}}\quad \Leftrightarrow\quad
(D-Id_{End(\hat{\bfC})})\ket{\Phi_{\infty}}=0\quad\leftrightarrow\quad
\left(\tfrac{d}{dt}\ket{\Psi(t)}\right)\big\vert_{t\to\infty}
=\left(H_U\ket{\Psi(t)}\right)\big\vert_{t\to\infty}=0
\end{equation}
However, CTMC simulations for rules that are not following mass-action semantics are non-standard to date, which is why it is also of interest to consider two particular mass-action type alternative CTMC models that are motivated by the original AVM model:
\begin{itemize}
\item A ``standard'' \emph{mass-action semantics (MAS)} CTMC, where the transformation rules in $D_{AVM}$ are defined to act in mass-action semantics (with the base rates ${\color{qOrange}\kappa_{\bullet}}, {\color{qOrange}\kappa_{\circ}},{\color{qOrange}\alpha_{\bullet}},{\color{qOrange}\alpha_{\circ}}\in \bR_{>0}$ however not fixed in any evident way from the original AVM model interpretation):
\begin{subequations}
\begin{align}
H_{MAS}&={\color{qOrange}\kappa_{\bullet}}\cdot
\rho\left(\delta\left(\begin{matrix}\bullet \hphantom {-}\circ \\[-0.675em] \!\!\!\!\!\!\backslash \\[-0.675em]\bullet \end{matrix}\;\leftharpoonup \;\begin{matrix}\bullet \!\!-\!\!\circ \\[-0.25em]\bullet \end{matrix}\right)\right)
+{\color{qOrange}\kappa_{\circ}}\cdot\rho\left(\delta\left(
\begin{matrix}\bullet \hphantom {-}\circ  \\[-0.675em] \;\;\;/\\[-0.675em]\circ \end{matrix}\;\leftharpoonup \;\begin{matrix}\bullet \!\!-\!\!\circ \\[-0.25em]\circ \end{matrix}\right)\right)\label{eq:AVM-H-MAS-rewire}\\
&\quad+{\color{qOrange}\alpha_{\bullet}}\cdot\rho\left(\delta\left(
\bullet \!\!-\!\!\bullet \;\leftharpoonup \;\bullet \!\!-\!\!\circ
\right)\right)
+{\color{qOrange}\alpha_{\circ}}\cdot\rho\left(\delta\left(
\circ \!\!-\!\!\circ \;\leftharpoonup \;\bullet \!\!-\!\!\circ
\right)\right) \label{eq:AVM-H-MAS-adopt}\\
&\quad-\hat{O}_{\bullet\!-\!\circ}({\color{qOrange}\kappa_{\bullet}}(\hat{O}_{\bullet}-1)+{\color{qOrange}\kappa_{\circ}}(\hat{O}_{\circ}-1)+{\color{qOrange}\alpha_{\bullet}}+{\color{qOrange}\alpha_{\circ}})\label{eq:AVM-H-MAS-diag}\,.
\end{align}
\end{subequations}
Note that if we let $\rho(h_{MAS}):=\eqref{eq:AVM-H-MAS-rewire}+\eqref{eq:AVM-H-MAS-adopt}$, then the terms in~\eqref{eq:AVM-H-MAS-diag} are indeed found to be equal to $-\jcOp{h_{AVM}}$ (utilizing~\eqref{eq:auxObsAVM} to simplify terms) as required by Theorem~\ref{def:masCTMC}.
\item A \emph{least-common-multiple (LCM)} variant which is computable from $D_{AVM}$ via the method presented in Example~\ref{ex:LCM} (with constant and operator-valued contributions to firing rates in {\color{qOrange}orange}):
\begin{subequations}
\begin{align}
H_{LCM}&=
{\color{qOrange}\frac{\alpha}{2N_E}}
\rho\left(\delta\left(\begin{matrix}\bullet \hphantom {-}\circ \\[-0.675em] \!\!\!\!\!\!\backslash \\[-0.675em]\bullet \end{matrix}\;\leftharpoonup \;\begin{matrix}\bullet \!\!-\!\!\circ \\[-0.25em]\bullet \end{matrix}\right)\right){\color{qOrange}(\hat{O}_{\circ}-1)}
+{\color{qOrange}\frac{\alpha}{2N_E}}\rho\left(\delta\left(\begin{matrix}\bullet \hphantom {-}\circ  \\[-0.675em] \;\;\;/\\[-0.675em]\circ \end{matrix}\;\leftharpoonup \;\begin{matrix}\bullet \!\!-\!\!\circ \\[-0.25em]\circ \end{matrix}\right)\right){\color{qOrange}(\hat{O}_{\bullet}-1)}\label{eq:H-AVM-LCM-rewire}\\
&\begin{aligned}
&\quad+{\color{qOrange}\frac{(1-\alpha)}{2N_E}}\rho\left(\delta\left(
\bullet \!\!-\!\!\bullet \;\leftharpoonup \;\bullet \!\!-\!\!\circ
\right)\right){\color{qOrange}(\hat{O}_{\bullet}-1)(\hat{O}_{\circ}-1)}\\
&\quad+{\color{qOrange}\frac{(1-\alpha)}{2N_E}}\rho\left(\delta\left(
\circ \!\!-\!\!\circ \;\leftharpoonup \;\bullet \!\!-\!\!\circ
\right)\right){\color{qOrange}(\hat{O}_{\bullet}-1)(\hat{O}_{\circ}-1)}\end{aligned} \label{eq:H-AVM-LCM-adopt}\\
&\quad-{\color{qOrange}\frac{1}{N_E}\hat{O}_{\bullet\!-\!\circ}{\color{qOrange}(\hat{O}_{\bullet}-1)(\hat{O}_{\circ}-1)}} \label{eq:H-AVM-LCM-diag}\,.
\end{align}
\end{subequations}
Expanding terms in $H_{LCM}$ via utilizing once again the representation property of $\rho$ (not presented here for brevity; cf.\ \cite[Thm.~3]{BK2020} for the details), we may write $H_{LCM}$ in the equivalent form below, which exhibits this CTMC model as a particular instance of a model for which all rules have the same input motif (and thus as a model in the spirit of the ``Potsdam approach'' as in~\cite{giese2012}):
\begin{subequations}\label{eq:potsdam-rules}
\begin{align} 
H_{LCM}&={\color{qOrange}\frac{\alpha}{2N_E}}\rho\left(\delta\left(
\begin{matrix}
\bullet \hphantom {-}\circ\\[-0.65em]
| \hphantom{-\circ}\\[-0.65em]
\bullet \hphantom {-}\circ
\end{matrix}\;\leftharpoonup\;
\begin{matrix}
\bullet \!\!-\!\!\circ\\[-0.65em]
\hphantom{-\circ}\\[-0.65em]
\bullet \hphantom {-}\circ
\end{matrix}
\right)\right)
+{\color{qOrange}\frac{\alpha}{2N_E}}\rho\left(\delta\left(
\begin{matrix}
\bullet \hphantom {-}\circ\\[-0.65em]
\hphantom{\bullet-}|\\[-0.65em]
\bullet \hphantom {-}\circ
\end{matrix}\;\leftharpoonup\;
\begin{matrix}
\bullet \!\!-\!\!\circ\\[-0.65em]
\hphantom{-\circ}\\[-0.65em]
\bullet \hphantom {-}\circ
\end{matrix}
\right)\right)\label{eq:H-AVM-LCM-alt-rewire}\\
&\quad+{\color{qOrange}\frac{1-\alpha}{2N_E}}\rho\left(\delta\left(
\begin{matrix}
\bullet \!\!-\!\!\bullet\\[-0.65em]
\hphantom{-\circ}\\[-0.65em]
\bullet \hphantom {-}\circ
\end{matrix}\;\leftharpoonup\;
\begin{matrix}
\bullet \!\!-\!\!\circ\\[-0.65em]
\hphantom{-\circ}\\[-0.65em]
\bullet \hphantom {-}\circ
\end{matrix}
\right)\right)
+{\color{qOrange}\frac{1-\alpha}{2N_E}}\rho\left(\delta\left(
\begin{matrix}
\circ \!\!-\!\!\circ\\[-0.65em]
\hphantom{-\circ}\\[-0.65em]
\bullet \hphantom {-}\circ
\end{matrix}\;\leftharpoonup\;
\begin{matrix}
\bullet \!\!-\!\!\circ\\[-0.65em]
\hphantom{-\circ}\\[-0.65em]
\bullet \hphantom {-}\circ
\end{matrix}
\right)\right)\label{eq:H-AVM-LCM-alt-adapt}\\
&\quad -{\color{qOrange}\frac{1}{N_E}\hat{O}_{\text{\tiny{$\begin{matrix}
\bullet \!\!-\!\!\circ\\[-0.65em]
\hphantom{-\circ}\\[-0.65em]
\bullet \hphantom {-}\circ
\end{matrix}$}}}}\label{eq:H-AVM-LCM-alt-diag}
\end{align}
\end{subequations}
\end{itemize}
\end{example}

\section{Simulating the Models}\label{sec:simulations}
In Section~\ref{sec:dtmc} we presented several ways of converting a DTMC-based model into a CTMC-based one. The first (M1) is based on introducing weight factors correcting for the deviation of the probabilistic view of the DTMC from the CTMC mass action semantics. This allows us to use a modified stochastic simulation algorithm to recover the behavior of the DTMC. Then we have the mass action stochastic graph transformation system (M2) as introduced in Section~\ref{sec:voter-models}, in our view the most natural and closest to reality due to its use of continuous time, but in order to relate to the original probabilistic formulation we have to sample rate constants experimentally until the behavior matches that observed in the given model. Finally we consider a model (M3) obtained by mutually extending the rules of the system to the same left-hand side. Then all rules have the same number of matches and hence rates directly reflect probabilities. 

In this section we simulate these models with the CTMC-based SimSG tool~\cite{DBLP:journals/jot/EhmesFS19} to see how, with suitably chosen parameters, their behavior matches that of the original formulation~\cite{Durrett2012}. In particular, we want to answer the following questions.
\begin{description}
\item[RQ1:] Can we reproduce the results of~\cite{Durrett2012} by simulating (M1) in SimSG?
\item[RQ2:] What rates do we have to use to reproduce the behavior of~\cite{Durrett2012} in (M2)?
\item[RQ3:] Does (M3) with rates reflecting the probabilities of~\cite{Durrett2012} lead to the same overall behavior?
\end{description}
Note that in (M1)  and (M2) rates are derived analytically based on the theory in Section~\ref{sec:dtmc} while they are determined experimentally for (M3). 
\begin{figure}[h] 
	\centering
	\subfigure[Initial graph (u=0.5)]{\includegraphics[width=0.31\textwidth]{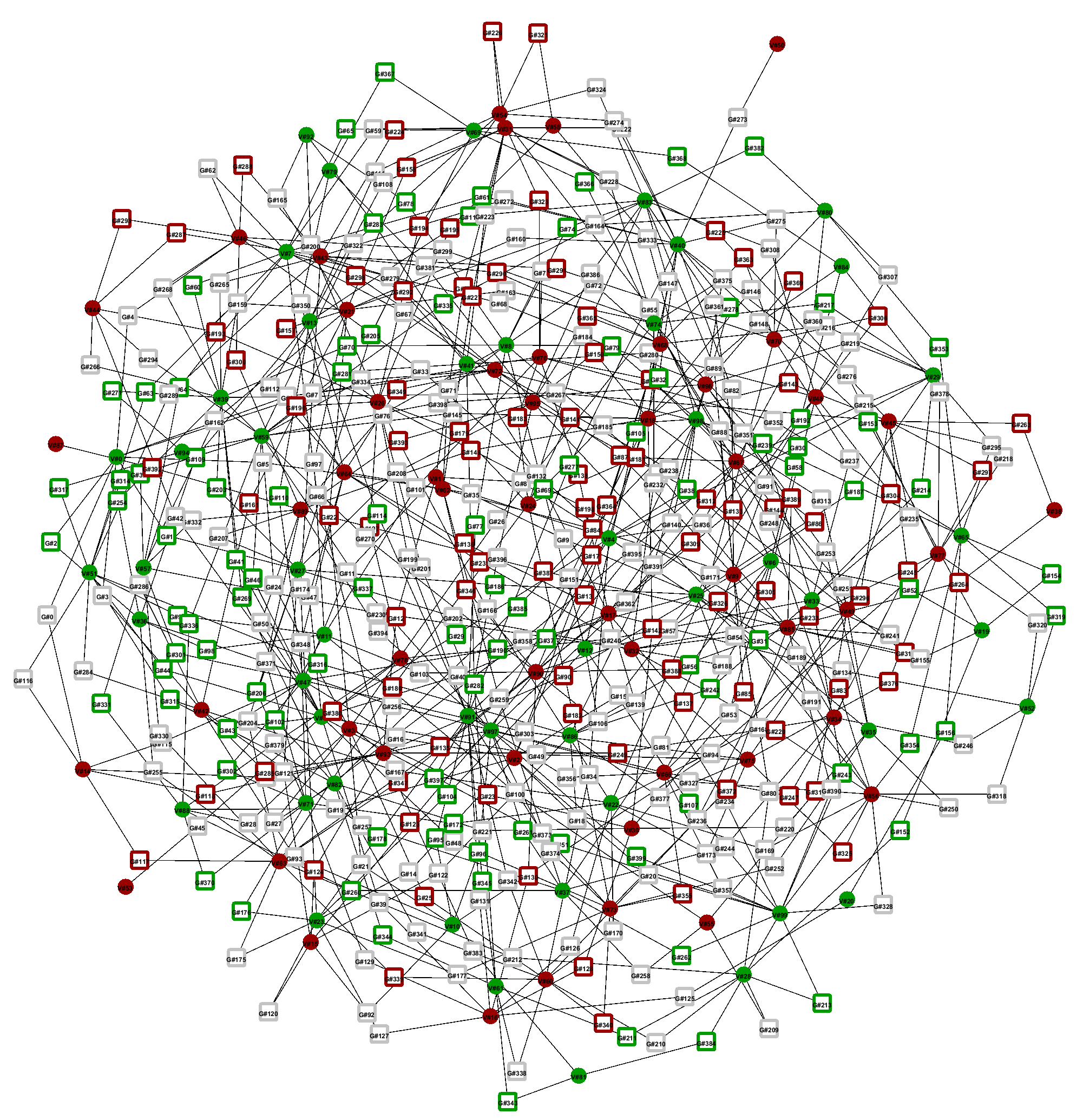}}
	\hfill
	\subfigure[Giant component]{\includegraphics[width=0.31\textwidth]{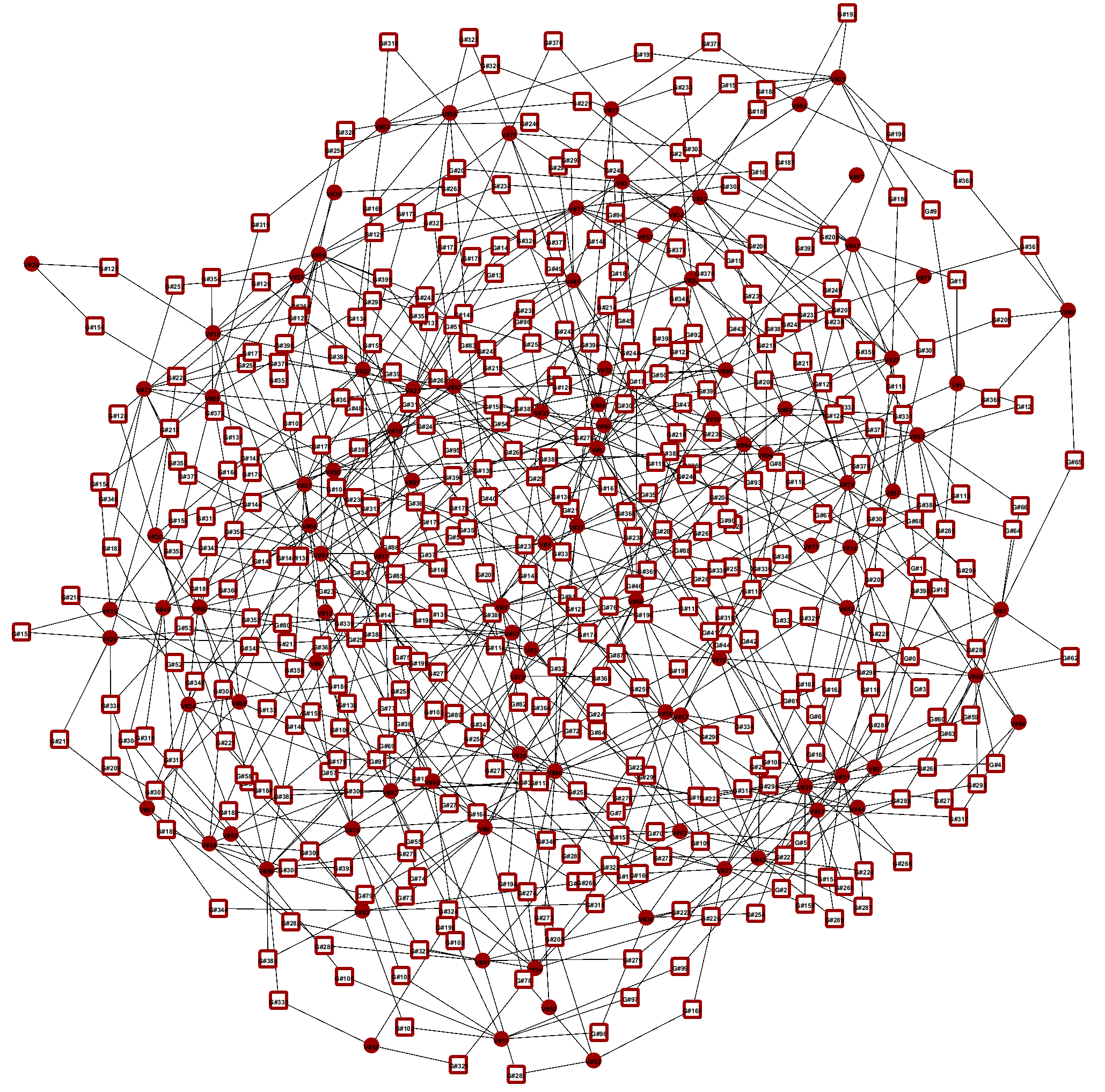}}
	\hfill
	\subfigure[Fragmented graph]{\includegraphics[width=0.31\textwidth]{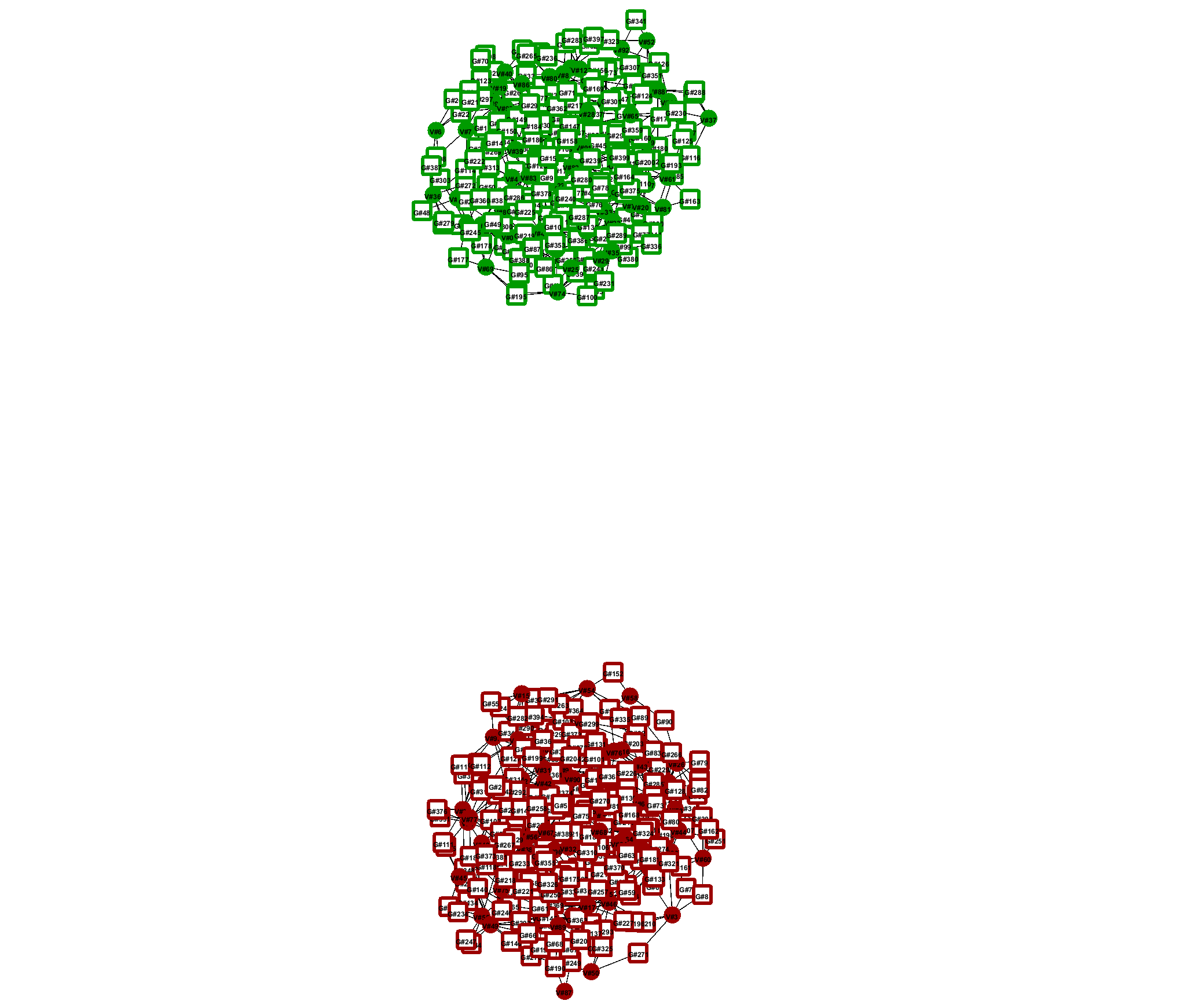}}
	\caption{Graphs before and after simulation}\label{fig:instance-graphs}
\end{figure}
As in~\cite{Durrett2012} initial graphs are generated randomly based on a fraction $u$ of agents voting 0 (where $u = 0.5$ gives a 50/50 split between opinions) by deciding for each pair of nodes with probability $v$ if they should be linked. This means that, with 100 voters and $v = 0.04$, we create an Erd{\"o}s-R{\'en}yi graph with 400 links and average degree 8.
According to~\cite{Durrett2012} we expect to see one of two behaviors, depending on the probability $\alpha$ of rewiring (vs. $1 - \alpha$ for adoption) for a given discordant link. With $\alpha > 0.43$ rewiring causes a fragmentation of the graph into homogeneous connected components; otherwise adoption leads to the dominance of a single opinion (usually the initial majority) in a giant component.  
Visually we indicate opinion 1 in green and opinion 0 in red. Homogeneous components are highlighted in their respective color, whereas heterogeneous components are shown in gray. For example, Figure~\ref{fig:instance-graphs}(a) shows the initial graph. In the limit, if the rates favor the adopt rules we obtain a graph as in Figure~\ref{fig:instance-graphs}(b), while Figure~\ref{fig:instance-graphs}(c) shows a possible result of dominant rewiring where the graph has split split into two components, one for each opinion.
\begin{figure}[h] 
	\centering
	\subfigure[DTMC Version of the Voter Model]{\includegraphics[width=0.48\textwidth]{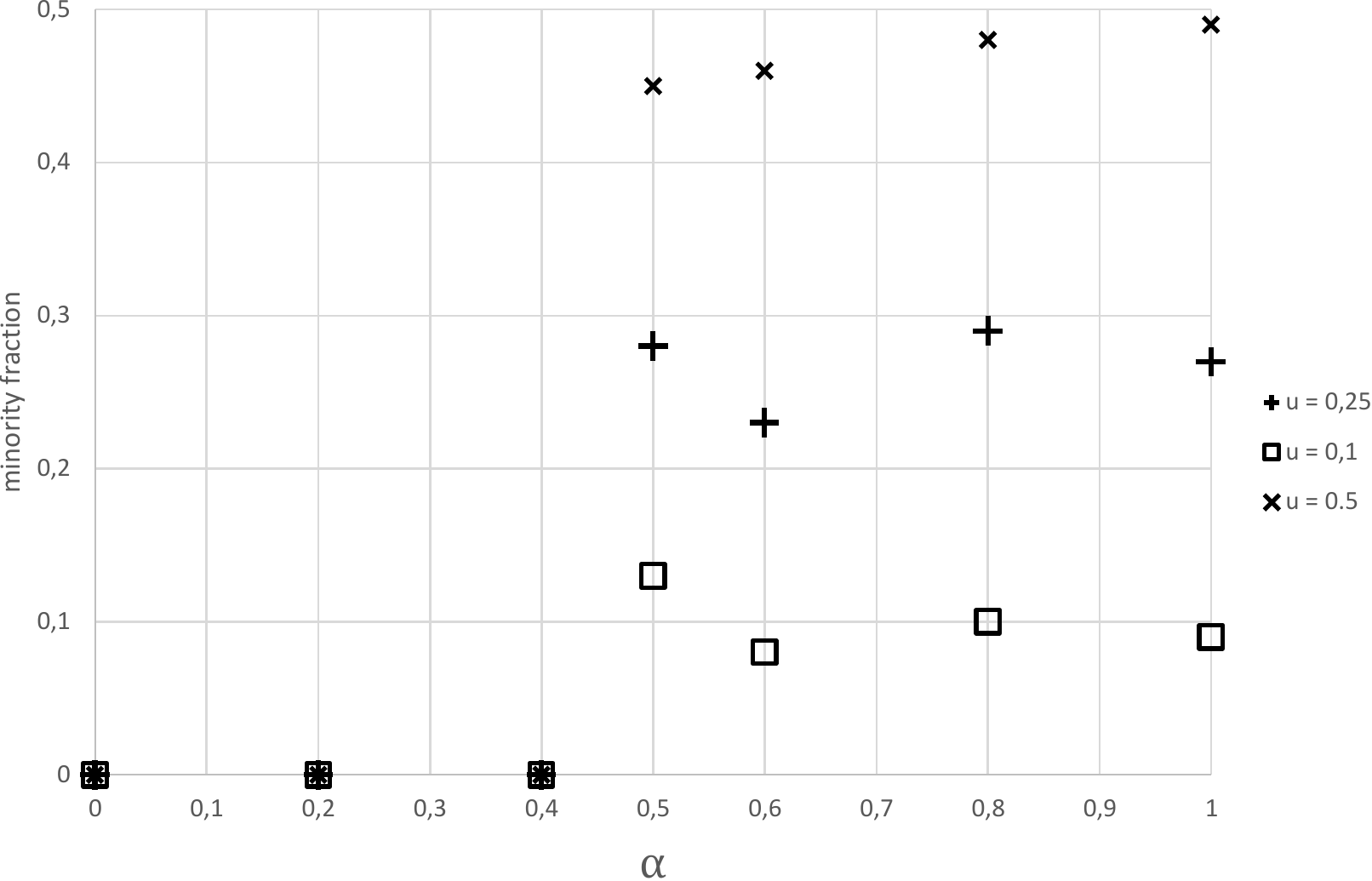}}
	\hfill
	\subfigure[CTMC Version of the Voter Model]{\includegraphics[width=0.48\textwidth]{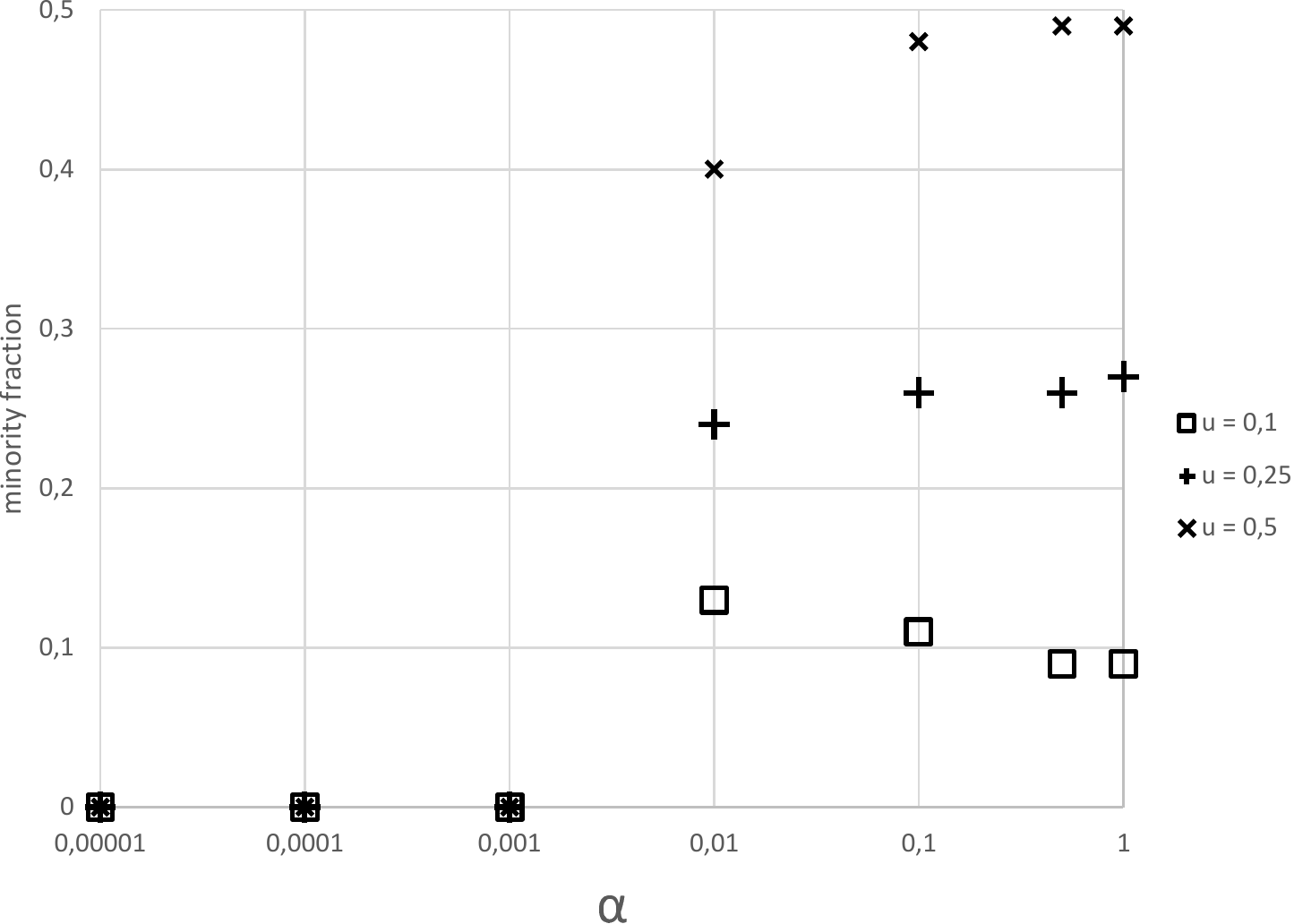}}
	\caption{Simulation results}\label{fig:sim-results}
\end{figure}

To answer RQ1 we use the rules as presented in Section~\ref{sec:voter-models}, but modify the simulation algorithm by the weight factors derived in Section~\ref{sec:dtmc}. This produces a simulation matching that of a discrete time Markov chain. While  fixing the model size, we perform a nested parameter sweep over $\alpha$, the static probability of rewiring, and over the fraction $u$ of agents voting $0$.
The results are shown in Figure~\ref{fig:sim-results}(a). On the $y$-axis we plot the fraction of voters with the minority opinion in the final graph, while the $x$-axis shows the \(\alpha\) value for each simulation. Furthermore, each set of symbols in the figure represents a different opinion ratio $u$ in the initial graph. As we can see, independently of the $u$ there is a clear point after which an increase in $\alpha$ leads to a separation of voters by opinion, which implies a segmentation of the graph. In turn, if $\alpha$ is low, the minority opinion disappears. Thus, Figure~\ref{fig:sim-results}(a) reflects very closely the findings of~\cite{Durrett2012}, with only a minor deviation of the value of $\alpha$ separating the two outcomes.	

For RQ2, we execute the same rules with our standard implementation simulation algorithm reflecting mass action CTMC semantics. While fixing the rate of the adopt rules as 1, we perform a nested parameter sweep over $\alpha$, in this case denoting the rate of the rewiring rules, and the minority fraction $u$. 
The results are shown in Figure~\ref{fig:sim-results}(b), again with the $y$-axis showing the fractions of voters in the minority in each final graph and the $x$-axis tracking the $\alpha$ values. As before, each set of symbols represents a different initial fraction $u$. In contrast to Figure~\ref{fig:sim-results}(a), here the threshold of $\alpha$ at which the final graph becomes segmented is orders of magnitudes smaller. This is a result of the CTMC-based simulation algorithm, that obtains the rule application probability by multiplying a rule's match count with its static rate. Intuitively, the set of matches for each of the adopt rules consists of all connected pairs of voters $v1, v2$ of different opinion. Instead, the set of matches of the rewire rules is made up by the Cartesian product of this first set with the set of voters sharing the opinion of $v1$.  To find the same balancing point as in the probabilistic model, this ``unfair advantage" of the rewire rules is compensated for by a very low rate.

\begin{figure}[h] 
	\centering
	\includegraphics[width=0.48\textwidth]{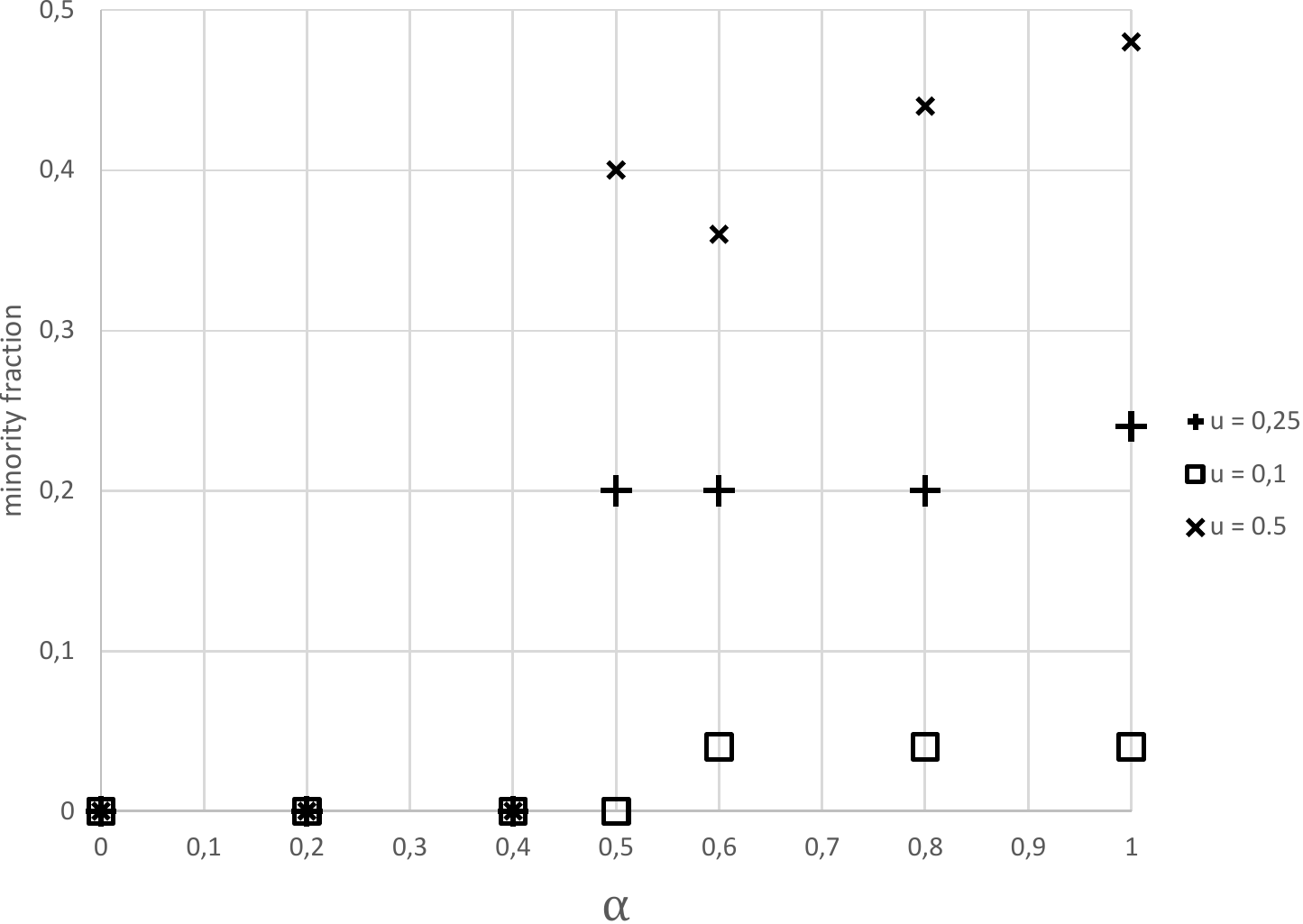}
	\caption{Simulation results of the alternative CTMC Version described by the rules in equation (\ref{eq:potsdam-rules})}\label{fig:sim-results2}
\end{figure}

Finally, to address RQ3 we perform simulations using adopt and rewire rules extended to a common left-hand side consisting of two connected voters $v1,v2$ of conflicting opinions plus one extra voter each of opinion $0$ and $1$. By the same argument as above this leads to a large increase in the number of matches and hence poses scalability challenges to the simulation. To address this we used a smaller initial graph of 50 Voters and 200 groups and executed only 10 simulation runs per parameter configuration rather than 40 as in the other two models. (In each case, configurations range across 3 values for $u$ and 7 values for $\alpha$ totaling 21 parameter configurations for each model.) The results are shown in Figure~\ref{fig:sim-results2} using the same layout as before.  
It is interesting that, despite the high degree of activity in this model due to the very large match count, it displays almost the same behavior as the DTMC version (M1), even down to the $\alpha$ threshold. 

The limited scalability of this approach is in part due to the global nature of the requirement that all rules share the same left-hand side. In the MDP-based approach of~\cite{giese2012} this applies only to the subsets of rules defining the same action. One could then consider the  four alternatives of rewire and adopt as cases of the same action of resolving a conflict edge, leading to a probabilistic graph transformation system with one rule. An analysis of the relation of mass-action CTMC with MDP-based semantics, however, is beyond the scope of this paper.

Social network analysis of the voter model and its variants is limited to a theoretical level, exploring mechanisms resulting in certain emergent phenomena that are observable in real-life networks, but are not quantified using real data. For a deeper quantitative analysis of phenomena such as the spread of opinions or the fragmentation of networks, the parameters of our rule-based models need to be matched to real social network data where such is available. This means, in particular, to determine the rates of the rules, e.g., from observations of pattern frequencies.
 
Such approaches are well established in chemical and biological modeling where statistical methods are used to derive rates of reaction rules from concentrations, i.e., relative pattern frequencies describing the ratios of the different types molecules~\cite{ Klinke2014InSM, Eydgahi2013MSB}. Instead, the derivation of an equational model from the operational rule-based description based on the rule algebra approach described in this paper could allow for an analytical approach where rates are determined by solving a system of equations. The precise conditions under which this is possible are a subject of future research.

\section{Conclusion}

In this paper we analyzed the non-standard semantics behind the formulation of adaptive system models in the literature. 
For a simple but prototypical example, we focused specifically on the voter models of opinion formation in social networks. We analyzed the non-standard semantics behind these models and established that they can be seen as Discrete-Time Markov Chains (DTMCs). In order to start the study of such models using the concepts, theories and tools of stochastic graph transformations (SGTS), we formalized the semantic relation between the rule-based specification by SGTS of mass-action Continuous-Time Markov Chains (CTMCs) with the DTMC-based semantics, identifying two systematic ways by which an SGTS can be derived from a DTMC-based model while preserving the behavior in the limit. 

In the first derivation, this leads to a generalized notion of SGTS with weight factors correcting for the mass-action component represented by the dependency of the jump rate on the number of matches for each rule. This new type of SGTS is supported by a generalized simulation algorithm supporting the analysis of DTMC-based probabilistic graph transformation systems which, to our knowledge, is original. The second derivation converts the rules of the system by extending them to the same left-hand sides, producing a model resembling Markov-Decision Process (MDP)-based probabilistic graph transformation rules. 

Apart from the theoretical analysis, we validate both resulting systems through simulations, establishing that they reproduce the expected behavior of the model. We also create a direct model of the same system as an SGTS with standard mass-action CTMC semantics and succeed in determining its parameters experimentally to match the expected behavior. We conclude that we can use standard mass-action stochastic GTS to model the phenomena expressed by the voter models in the literature, providing a starting point for further more elaborate modeling and analyses. 

In future work we want to study more complex adaptive networks, including variations on the voter model with groups of more than two members, opinion profiles for a set of topics instead of a single opinion per voter, and including concepts such as influencers (who actively try to link to and persuade others), or zealots (who do not change their opinions). We are also planning to study how to match models to  social network data.
On the foundational side, we are planning optimizations to the simulation algorithm and a study of the relation between mass-action CTMC and MDP semantics of probabilistic graph transformations. We can also derive and study differential equations (ODEs) from our SGTS using the rule-algebra formalism. This provides a more scalable solution to analyzing their behavior complementing the simulation-based approach.

\end{document}